\documentclass[aps,prd,two column]{revtex4-2}

\usepackage{amsmath}
\usepackage{ulem}
\usepackage{color}
\usepackage{graphicx}
\usepackage{epsfig}
\usepackage{bm}
\usepackage{hyperref}
\usepackage{natbib}
\usepackage{caption}
\usepackage{subcaption}

\def\be {\begin{equation}}
\def\ee {\end{equation}}
\def\nn {\nonumber}
\def\bea {\begin{eqnarray}}
\def\eea {\end{eqnarray}}

\newcommand{\om}{\omega} 

\newcommand{\vk}{\vec k}

\newcommand{\del}{\partial}

\begin{document}

\title{Thermoelectric Thomson coefficient of quark-gluon plasma in the presence of a time-varying magnetic field }

\author{Kamaljeet Singh}
\email{kspaink84@gmail.com}
\author{Raghunath Sahoo}
\email{Corresponding Author: Raghunath.Sahoo@cern.ch}
\affiliation{ Department of Physics, Indian Institute of Technology Indore, Simrol, Indore 453552, India} 

\date{\today}
\begin{abstract}
Heavy-ion collision experiments such as the Large Hadron Collider and the Relativistic Heavy Ion Collider offer a unique platform to study several key properties of the quark-gluon plasma (QGP), a deconfined state of strongly interacting matter. Quarks, being the electrically charged particles, can induce an electric current in the medium in response to the temperature gradients. Hence, the QGP medium can behave like a thermoelectric medium. The thermoelectric coefficients, such as the Seebeck and Thomson coefficients, can help us to understand the intricate transport phenomenon of the medium. In peripheral collisions, the intense, transient, and time-dependent magnetic field created due to spectator protons significantly influences the thermoelectric properties of the QGP medium, affecting the charge and heat transport. This work uses the quasi-particle model to calculate the Thomson coefficient in QGP.  The Thomson effect, describing the continuous heating or cooling of the charge-carrying medium in the presence of temperature gradients, remains largely unexplored in QGP. The Seebeck effect, which relates temperature gradients to induced electric fields, has been widely studied in the literature.  For the first time, we calculate the magneto-Thomson and transverse Thomson coefficients. We have studied their dependence on temperature, baryon chemical potential, center of mass energy, and time-dependent magnetic field with different decay parameters. The transverse Thomson effect originates due to the presence of the Nernst effect in the presence of a magnetic field. Our results provide new insights into the higher-order thermoelectric transport properties of the QGP medium in the context of heavy-ion collisions. 
\end{abstract}
\maketitle

\section{Introduction}
The study of quark-gluon plasma (QGP) represents one of the most exciting frontiers in theoretical and experimental high-energy physics. QGP is a state of matter in which quarks and gluons, ordinarily confined within hadrons, become deconfined locally and create a strongly coupled medium \cite{Brewer:2019oha, Busza:2018rrf}. This state can possibly be created in relativistic heavy-ion collisions at facilities such as the Relativistic Heavy Ion Collider (RHIC) and the Large Hadron Collider (LHC), and its study deepens our understanding of quantum chromodynamics (QCD) matter under extreme conditions of temperature and energy density. The experimental signatures reveal that the QGP medium is a low viscosity medium often called a nearly perfect fluid~\cite{Bernhard:2019bmu}. These experiments also provide critical insights into the thermodynamic~\cite{Pradhan:2023rvf,Khaidukov:2019icg, Koothottil:2018akg, Goswami:2023eol,Sahoo:2023vkw} and transport properties~\cite{Gavin:1985ph, K:2022pzc} of QCD matter. Transport properties of the QGP medium are essential for understanding how energy, momentum, and conserved charges (such as baryon number and electric charge) are transferred within this strongly interacting medium. One can get a long list of references~\cite{Dey:2020awu,Pradhan:2022gbm,Huang:2011dc,Ghosh:2019ubc,Nam:2012sg,Hattori:2016cnt,Hattori:2016lqx,Harutyunyan:2016rxm,Kerbikov:2014ofa,Feng:2017tsh,Wang:2020qpx,Li:2019bgc,Tuchin:2012mf,Huang:2009ue,Agasian:2011st,Ghosh:2018cxb,Nam:2013fpa} for microscopic calculations of various transport coefficients, such as electrical conductivity, thermal conductivity, and shear viscosity, for the hot and dense QCD matter. The transport properties in condensed matter systems are dominated by one type of charge carrier, either electrons or holes. Therefore, the presence of temperature gradients leads to the thermoelectric transport in the system. On the other hand, the electron-ion plasma is a medium that has electrons and ions, both of which are mobile carriers. The different mobility of electrons and ions can give rise to thermoelectric effects in the presence of temperature gradients. In such a medium, the only temperature gradient can not ensure the thermoelectric effect due to the cancellation of the electric current of electrons and ions.

The thermoelectric transport of the QGP arises because its constituent quarks are electrically charged. As a temperature gradient develops in the plasma, provided a non-zero baryon chemical potential, it drives a non-zero net electric current, which manifests the thermoelectric effect~\cite{Zheng}. In condensed matter systems, the study of thermoelectric transport coefficients, such as Seebeck~\cite{PhysRevB.105.235116, PhysRev.98.940} and Thomson coefficients~\cite{PhysRevLett.125.106601, Image}, provides a sophisticated tool for understanding the complex transportation of heat and electric charge in the system. Analogous to the Seebeck effect in semiconductors, where a temperature gradient induces a voltage across the junction, the QGP medium can also show a Seebeck effect associated with conserved baryon current generated by temperature gradients~\cite{Singh:2024emy}. The temperature dependence of the Seebeck coefficient further introduces the higher-order thermoelectric effects such as the Thomson effect~\cite{Singh:2025rwc}. The Thomson coefficient quantifies the reversible heat exchanged per unit baryon-charge in the presence of a temperature gradient. The study of transport properties of QGP medium is intricate because of the presence of multiple degrees of freedom of the medium constituents. Microscopic investigations of the transport properties of quark and hadronic matter can be understood by following two major theoretical frameworks. On the one hand, the Kubo formalism based on quantum field theory~\cite{Harutyunyan:2017ttz, Fernandez-Fraile:2009eug} and on the other hand, the relaxation time approximation (RTA) to solve Boltzmann transport equation (BTE) within the kinetic theory framework~\cite{Gavin:1985ph, Das:2019pqd, Dash:2020vxk, K:2022pzc, Singh:2023pwf, Singh:2023ues, Singh:2024emy}. In previous studies on the thermoelectric properties of QGP medium, the Seebeck, magneto-Seebeck, and Nernst coefficients have been extensively investigated in the presence of a magnetic field.
~\cite{Abhishek:2020wjm, Kurian:2021zyb,K:2022pzc, Zhang:2020efz}. This is the first time we have looked into the detailed study of Thomson, the magneto-Thomson and transverse Thomson coefficient of the QGP medium.
In non-central heavy-ion collisions, the spectators lead to a strong electromagnetic field that magnetizes the QGP~\cite{Wang:2021eud, McInnes:2016dwk}. This generation of strong, transient electromagnetic fields can be predicted by classical electrodynamics~\cite{Tuchin:2015oka, Satow:2014lia, ALICE:2019sgg, Kharzeev:2007jp, STAR:2023jdd}. These fields, along with variations in collision energy and centrality, affect the equation of state (EoS) of the medium~\cite{Kumar:2024coh, Brandt:2022hwy, Borsanyi:2012cr, Endrodi:2011gv} and can drive several phenomena such as the chiral magnetic effect~\cite{Kharzeev:2007jp, Li:2017tgi} and magnetic/inverse magnetic catalysis~\cite{Gusynin:1994re, Lee:1997zj}, as well as transitions near the QCD critical point~\cite{Avancini:2012ee, Endrodi:2011gv}. These fields can reach magnitudes of 10-15 $m_\pi^2$ ($\approx 10^{18}$ Gauss), which is much higher than the QCD scale ($\Lambda_{QCD}\approx 1.5~m_\pi$), with $m_\pi \approx$140 MeV)\cite{Simonov:2021eyt}

This work focuses on a comprehensive study of the estimation of the thermoelectric Thomson and magneto-Thomson coefficients in the QGP medium. For the completeness of the work, we have also calculated the Seebeck and magneto-Seebeck coefficients. 
The paper is organized as follows. Sec.~\ref{sec:formalism} is dedicated to the detailed derivation of the Thomson coefficient by solving the Boltzmann transport equation (BTE) under relaxation time approximation (RTA). In this section, we have two subsections: subsection~ \ref{formalism1} is for the case of zero magnetic fields, and subsection~ \ref{formalism2} is for the case of non-zero magnetic fields, where we incorporate the effect of the external time-varying magnetic field and calculate the magneto-Thomson coefficient and transverse Thomson coefficient. In Sec.~\ref{Sec-results}, we discuss our obtained results for the calculations of thermoelectric coefficients following the formalism. Finally, Sec.~\ref{sec-summary} summarizes our work with possible outlooks.

\section{Formalism}\label{sec:formalism}
In this section, we calculate the Thomson coefficient for the QGP medium created in heavy-ion collisions for the cases of zero and non-zero magnetic fields. 
\subsection{Thomson coefficient for the case of zero magnetic field.}
\label{formalism1}
 To study the Thomson effect of QGP medium for the case of zero magnetic fields, we first consider the Boltzmann transport equation under relaxation time approximation. The RTA provides the framework where the Boltzmann equation can be interpreted as a linear expansion of the single-particle total distribution function ($f_i$) around the single-particle equilibrium distribution function ($f_i^{0}$). The $f_i$ can be written as $f_i=f^0_i+\delta f_i$, provided $\delta f_i$ represents the deviation from the equilibrium state. The total single-particle distribution function for $i^{\rm th}$ species at equilibrium is given by
\begin{align}\label{Dis-f}
f^0_i= \frac{1}{e^{\frac{\omega_i-{\rm b}_i\mu_B}{T}}\pm 1}~,
\end{align}
where $\omega_i=\sqrt{\vec{k_i}^2+m_i^2}$ is the single particle energy,  $\mu_B$ is the baryon chemical potential, $b_i$ denotes the baryon number of $i^{\rm th}$ species, e.g. for quarks $b_i=1/3$, for anti-quarks $b_i=-1/3$ and for gluons $b_i=0$. The $\pm$ sign stands for fermions and bosons, respectively. The linearized BTE under RTA in the local rest frame (LRF), for particle species $i$ can be written as \cite{Das:2020beh, Singh:2023pwf},

\begin{equation}
\frac{\partial f_i}{\partial t} + \vec{v}_i.\vec{\nabla}f_i+q_i\vec{E}.\frac{\partial f_i}{\partial\vec{k_i}} = 
  -\frac{\delta f_i(\vec{x_i},\vec{k_i})}{\tau^{i}_R} ,
 \label{equnew1}
\end{equation}
where $\vec{E}$ is the non-zero electric field that drives the system out of thermal equilibrium and  $\tau^{i}_R$ denotes the relaxation time of the particle species $i$. The equilibrium distribution function satisfies, 
\begin{align}
 \frac{\partial f^{0}_{i}}{\partial \vec{k_i}}=\vec{v}_i\frac{\partial f^{0}_{i}}{\partial \omega_i},
 ~~ \frac{\partial f^{0}_{i}}{\partial \omega_i}=-\frac{f^{0}_{i}(1\mp f^{0}_{i})}{T}, 
\label{equnew2}
 \end{align}
  $\vec{v}_i=\vec{k_i}/\omega_i$ is the velocity of the particle. For a system having local thermodynamic equilibrium, the spatial dependence of the distribution functions appears due to spatial gradients of temperature and baryon chemical potential. The gradient of the equilibrium distribution function $\vec{\nabla}f^{0}_{i}$ can be expressed as,
\begin{align}
 \vec{\nabla}f^{0}_{i} = T \bigg[\omega_i\vec{\nabla}\left(\frac{1}{T}\right)-b_i\vec{\nabla}\left(\frac{\mu_B}{T}\right)\bigg]\frac{\partial f^{0}_{i}}{\partial\omega_i}.
 \label{equnew3}
\end{align}

Using the Gibbs-Duhem relation, we then have,
\begin{align}
 \vec{\nabla}f^{0}_{i} = - \frac{\partial f^{0}_{i}}{\partial\omega_i}\bigg(\omega_i-{\rm b}_i \xi\bigg)\frac{\vec{\nabla}T}{T}.
\label{equnew4}
 \end{align}
 where $\xi=\frac{\varepsilon + P}{n}$ is the enthalpy per particle, $\varepsilon$, $P$, and $n$ are total energy density, total pressure, and net baryon density of the system, respectively.
 With leading order contributions, we can write an ansatz of $\delta f_i$ as~\cite{Gavin:1985ph, Singh:2023pwf}
\begin{align}\label{delta-f0}
	\delta f_i = (\vec{k_i} \cdot \vec{\Omega}) \frac{\partial f^0_i}{\partial \omega_i}~.
\end{align}  
In general, the preferred form of unknown vector ${\vec \Omega}$ can be assumed as a linear combination of all existing perturbing forces leading the system out of thermal equilibrium as \begin{align}\label{Omega00}
\vec{\Omega} = &~\alpha_1 \vec{E} +\alpha_2 \vec{\nabla}T ~.
\end{align}  
The unknown coefficients $\alpha_j$ ($j=1,2$) determine the strength of the respective gradient force fields driving the system away from equilibrium.
Using Eq.\eqref{equnew4} and Eq.\eqref{equnew2} in Eq.\eqref{equnew1}, we can write the deviation of the equilibrium distribution function as,
\begin{align}
 \delta f_i = -\tau^{i}_R\frac{\partial f_i^{0}}{\partial \omega_i}\bigg[q_i(\vec{E}.\vec{v}_i)-\bigg(\frac{\omega_i-{\rm b}_i \xi}{T}\bigg)\vec{v}_i.\vec{\nabla} T\bigg].
 \label{equnew5}
\end{align}
Following the kinetic theory, the electric current $(\vec{j})$ of the system can be written in terms of the deviation from the equilibrium distribution function $\delta f_i$ as,
\begin{align}
 \vec{j}& = \sum_i g_i\int \frac{d^3|\vk_i|}{(2\pi)^3}q_i\vec{v}_i\delta f_i\nonumber\\
 & = \sum_i \frac{g_i}{3} \int \frac{d^3|\vk_i|}{(2\pi)^3}\tau^{i}_R q_i^2 v^2_i\bigg(-\frac{\partial f_i^{0}}{\partial \omega_i}\bigg)\vec{E}\nonumber\\
 & -\sum_i \frac{g_i}{3} \int \frac{d^3|\vk_i|}{(2\pi)^3}\tau^{i}_Rq_iv^2_i
 \bigg(\frac{\omega_i-{\rm b}_i \xi}{T}\bigg)\bigg(-\frac{\partial f_i^{0}}{\partial \omega_i}\bigg)\vec{\nabla} T.
 \label{equnew6}
\end{align}
In the above equation, we have used $\langle v^l_iv^j_i\rangle=\frac{1}{3}v_i^2\delta^{lj}$. Here, the sum is over all the quarks and anti-quarks. 
For a relativistic system, one can also define the thermal current with reference to the conserved baryon current. The thermal current arises when energy flows relative to the baryonic enthalpy. 
Hence, the heat current of the QGP medium can be defined as \cite{Gavin:1985ph},

\begin{align}
\label{heat_current}
 \vec{I} & = \sum_i g_i\int \frac{d^3|\vk_i|}{(2\pi)^3}k_if_i - \xi\sum_i b_i g_i\int\frac{d^3|\vk_i|}{(2\pi)^3} v_i f_i\nonumber\\
 & = \sum_i g_i\int \frac{d^3|\vk_i|}{(2\pi)^3}\frac{k_i}{\omega_i}\left(\omega_i-{\rm b}_i \xi\right)\delta f_i.
 \end{align}
 Substitute the value of $\delta f_i$ from Eq.~(\ref{equnew5}) into Eq.~(\ref{heat_current})
 \begin{align}
 \vec{I}& = \sum_i \frac{g_i}{3} \int \frac{d^3|\vk_i|}{(2\pi)^3} \tau^{i}_R q_i v_i^2\left(\omega_i-{\rm b}_i \xi\right)\left(-\frac{\partial f_i^{0}}{\partial\omega_i}\right)\vec{E}\nonumber\\
 & -\sum_i \frac{g_i}{3T} \int \frac{d^3|\vk_i|}{(2\pi)^3} \tau^{i}_R v_i^2\left(\omega_i-{\rm b}_i \xi\right)^2\left(-\frac{\partial f_i^{0}}{\partial\omega_i}\right)\vec{\nabla}T.
 \label{equnew9}
\end{align}
One can define the Seebeck coefficient $S$ using Eq.\eqref{equnew6} by setting $\vec{j}=0$ such that the electric field and temperature gradient become proportional to each other. Here, the proportionality factor is known as the Seebeck coefficient \cite{Singh:2024emy}. Hence from  Eq.\eqref{equnew6} we get, 
\begin{align}
 \vec{E}=S\vec{\nabla}T,
\end{align}
hence, 
\begin{align}
 S &= \frac{\sum_i \frac{g_i}{3}\int \frac{d^3|\vk_i|}{(2\pi)^3}\tau^{i}_R q_i v_i^2\left(\omega_i-{\rm b}_i \xi\right)\left(-\frac{\partial f_i^{0}}{\partial\omega_i}\right)}{T\sum_i \frac{g_i}{3}\int \frac{d^3|\vk_i|}{(2\pi)^3}\tau^{i}_R q^2_i v_i^2\left(-\frac{\partial f_i^{0}}{\partial\omega_i}\right)}\nonumber\\
 &=\frac{\sum_i \frac{g_i}{3T}\int \frac{d^3|\vk_i|}{(2\pi)^3}\tau^{i}_R q_i \left(\frac{\vec{k_i}}{\omega_i}\right)^2\left(\omega_i-{\rm b}_i \xi\right)f^{0}_{i}(1\mp f^{0}_{i})}{T\sum_i \frac{g_i}{3T}\int \frac{d^3|\vk_i|}{(2\pi)^3}\tau^{i}_R q^2_i \left(\frac{\vec{k_i}}{\omega_i}\right)^2f^{0}_{i}(1\mp f^{0}_{i})}\nonumber\\
 &= \frac{\mathcal{I}_{1}/T^2}{\sigma_{el}/T}.
 \label{equnew11}
\end{align}
Where, the electrical conductivity ($\sigma_{el}$) can be identified from Eq.\eqref{equnew6} as,
\begin{align}
 \sigma_{el}  & = \sum_i \frac{g_i}{3T}\int \frac{d^3|\vk_i|}{(2\pi)^3}\tau^{i}_R q^2_i \left(\frac{\vec{k_i}}{\omega_i}\right)^2f^{0}_{i}(1\mp f^{0}_{i}),
\end{align}
and the integral $\mathcal{I}_{1}$ in Eq.\eqref{equnew11} is, 
\begin{align}
 \mathcal{I}_{1} = \sum_i \frac{g_i}{3T}\int \frac{d^3|\vk_i|}{(2\pi)^3}\tau^{i}_R q_i \left(\frac{\vec{k_i}}{\omega_i}\right)^2\left(\omega_i-{\rm b}_i \xi\right)f^{0}_{i}(1\mp f^{0}_{i}). 
 \label{I31equ}
\end{align}
It is to be noted that the Seebeck coefficient can be both positive and negative because the numerator depends linearly on an electric charge while the integrand itself is not positive definite. As the Seebeck coefficient $S$ for the QGP medium in Eq.~(\ref{equnew11}) is temperature dependent, the Thomson effect originates in the medium. The Thomson coefficient ($Th$) is related to the Seebeck coefficient as
\begin{align}\label{Thomson}
    Th = T\frac{dS}{dT}.
\end{align}
The above relation is usually known as the first Thomson relation, and it can be derived from energy conservation~\cite{PhysRevLett.125.106601}. The electric current and heat current can be modified due to these thermoelectric coefficients as,
\begin{align}
 \vec{j}&=\sigma_{el}\vec{E}-\sigma_{el}S \vec{\nabla}T. \label{equnew14a}\\
 \vec{{I}}&=T\sigma_{el}S\vec{E}-\kappa_0\vec{\nabla}T,
 \label{equnew14}
\end{align}
where $\kappa_0$ is the coefficient of the thermal conductivity and is expressed as \cite{Singh:2023pwf},
\begin{align}
 \kappa_0=\sum_i\frac{g_i}{3T^2}\int\frac{d^3|\vk_i|}{(2\pi)^3}\tau^{i}_R\left(\frac{\vec{k_i}}{\omega_i}\right)^2\left(\omega_i-{\rm b}_i \xi\right)^2f^{0}_{i}(1\mp f^{0}_{i}).
 \label{equnew15}
\end{align}
Using Eq.\eqref{equnew14a} and Eq.\eqref{equnew14}, we can express the heat current $\vec{{I}}$ in terms  of electric current $\vec{j}$ in the following way,
\begin{equation}
 \vec{{I}}=TS\vec{j}-\left(\kappa_0-T\sigma_{el}S^2\right)\vec{\nabla}T.
 \label{equnew16}
\end{equation}
 It is to be noted that gluons contribute through the total enthalpy of the medium only as they have zero electric charge; hence, their contribution enters through the numerator of Eq.\eqref{equnew11}. One can use the Eqs~(\ref{equnew14}),~(\ref{equnew15}) to write the local energy balance equation for a thermoelectric medium as~\cite{Chiba_2023}
 \begin{align}
     \vec{\nabla}(-\kappa_0\vec{\nabla}T) = \frac{(\vec{j})^2}{\sigma_{el}} - Th\vec{\nabla}Tj 
 \end{align}
 The term on the left side represents the heat conducted through the medium due to the temperature gradient. It follows Fourier’s Law of heat conduction, which states that heat flows from regions of high temperature to low temperature. Whereas the first term on the right side of the equation represents the heat generated due to electrical resistance ($1/\sigma_{el}$) as current flows through the medium. The second term describes the Thomson effect, which occurs when there is both an electric current and a temperature gradient. As the Thomson term is proportional to $\vec{j}$, depending on the direction of the current and the temperature gradient, heat can be absorbed or released. Hence, the presence of the Thomson coefficient may significantly affect the cooling of the medium. The Thomson coefficient is non-zero when a particular medium has a temperature-dependent Seebeck coefficient. 
 The presence of a magnetic field makes this picture more complicated, which we discuss in the next section.

\subsection{Magneto-Thomson coefficient for the case of non-zero magnetic field.}
\label{formalism2}
The BTE for a single particle species under RTA in the presence of an external electromagnetic field can be expressed as, 
\begin{align}
 \frac{\partial f_i}{\partial t} +\vec{v}_i.\frac{\partial f_i}{\partial \vec{x_i}}+q_i\left(\vec{E}+\vec{v}_i\times\vec{B}\right).\frac{\partial f_i}{\partial\vec{k_i}} = -\frac{\delta f_i}{\tau^{i}_R},
 \label{equnew21}
\end{align}
Here, we consider a time-varying electromagnetic field of the form ~\cite{Satow:2014lia, Hongo:2013cqa}
\begin{align}
	B = B_0 \exp{\left(-\frac{\tau}{\tau_B}\right)},~~&~~
	E = E_0 \exp{\left(-\frac{\tau}{\tau_E}\right)},\label{Mag-Field}
\end{align}
where $B_0,~ E_0$ are the magnitudes of the initial fields having decay parameters of $\tau_B$ and $\tau_E$, respectively, and $\tau$ is the proper time. The exponential decay forms for the electric and magnetic fields are motivated by solutions to Maxwell’s equations in a conducting medium, where finite electrical conductivity leads to time-dependent damping of the fields. In particular, such forms arise naturally when Ohm’s law is coupled to Maxwell’s equations. For simplicity, we focus on their time evolution in proper time $\tau$, suppressing spatial dependence, which is assumed to be smooth.  
To solve the Eq.~(\ref{equnew21}), we take an ansatz to express the deviation of the distribution function from the equilibrium in the following way~\cite{Gavin:1985ph,Das:2021qii},
  \begin{align}
  \delta f_i = (\vec{k_i}.~\vec{\Omega})\frac{\partial f^{0}_{i}}{\partial\omega_i},
\label{equnew22}
  \end{align}
with $\vec{\Omega}$ being related to a temperature gradient, electric field, the magnetic field, and in general, can be written as,
\begin{align}\label{equnew23}
\vec{\Omega} = &~\alpha_1 \vec{E} + \alpha_2 \dot{\vec E} +\alpha_3 \vec{\nabla}T + \alpha_4 (\vec{\nabla}T \times \vec{B}) 
 + \alpha_5 (\vec{\nabla}T\times \dot{\vec{B}})\nn\\
+ &\alpha_6 (\vec{E} \times {\vec{B}}) 
 + \alpha_7 (\vec{E} \times {\dot{\vec{B}}})
 + \alpha_{8}(\dot{\vec{E}} \times {\vec{B}}),
\end{align}  
where the unknown coefficients $\alpha_j$ ($j=1,2,..8$) can determine the strength of the respective field in driving the system away from the equilibrium. In the above equation, we have considered all the forces responsible for driving the system away from equilibrium according to their consistency with CP symmetries of electric and heat current. The external time-varying electromagnetic field, temperature gradient, and their cross terms up to leading order are considered here. For the time-independent electromagnetic field the coefficients $\alpha_2$, $\alpha_5$, $\alpha_7$, and $\alpha_{8}$ vanish. For the case where the chiral chemical potential is zero, the terms $\vec{B}$, $\dot{\vec B}$ do not contribute to the current~\cite{Satow:2014lia}. 

After getting the expressions for $\alpha_j$'s, the simplified form of $\delta f_i$ is (see Appendix of Ref.~\cite{Singh:2024emy}),
\begin{align}
 \delta f_i &= \frac{-q_i \tau_R^i}{(1+\chi_i)(1+ \chi_i + \chi_i^2)}\Big[\Big\{(1+\chi_i)+\frac{\chi_i(1+ \chi_i - \chi_i^2)}{(1+ \chi_i^2)}\Big\}\nn\\
 &(\vec{v_i}.\vec{E}) \pm \Big\{\chi_i(1+\chi_i)
 + \frac{\chi_i^2(2+ \chi_i)}{(1+ \chi_i^2)}\Big\}\Big(\vec{v_i}.(\vec{E}\times \hat{b})\Big)\Big]\nn\\ 
 &+ \frac{(\omega_i-{\rm b}_i \xi) \tau_R^i}{T(1+\chi_i)(1+ \chi_i + \chi_i^2)}\Big[(1+ \chi_i)\Big(\vec{v_i}\cdot\vec{\nabla}{T}\Big)\nn\\
 &\pm\chi_i(1+\chi_i)
 \Big(\vec{v_i}\cdot(\vec{\nabla}{T} \times \hat{b})\Big)
 \Big]\frac{\partial f^0_i}{\partial \om_i}.
\end{align}
For simplicity we have considered $\tau_E = \tau_B$ and $\chi_i = \frac{\tau_R^i}{\tau_B} = \frac{\tau_R^i}{\tau_E}$. 
Here, $\pm$ indicates a positive sign for positively charged particles (antiparticles) and a negative sign for negatively charged particles (antiparticles).
Now, we can express the electrical current using $\delta f_i$ as,
\begin{align}\label{equnew36}
j^l &= \sum_{i} \frac{q_i g_i}{3} \int \frac{d^3|\vk_i|}{(2\pi)^3} v_i^2 \frac{\tau_R^i}{(1+ \chi_i + \chi_i^2)(1+\chi_i)}\nn\\
&\Big[-q_i\Big\{ \Big(
(1+\chi_i)
+\frac{\chi_i(1+ \chi_i - \chi_i^2)}{(1+ \chi_i^2)}\Big)\delta^{lk}E^k \nn\\
&\pm \Big(\chi_i(1+\chi_i)+
\frac{\chi_i^2(2+ \chi_i)}{(1+ \chi_i^2)} \Big)
\epsilon^{ljk}h^jE^k\Big\}\nn\\
&+ \frac{(\omega_i-{\rm b}_i \xi)}{T}\Big\{(1+\chi_i)\delta^{lk}
\frac{\del {T}}{\del x^k}\nn\\
&\pm\chi_i(1+\chi_i)\epsilon^{ljk}h^j
\frac{\del {T}}{\del x^k}\Big\} \Big]\frac{\partial f^0_i}{\partial \om_i}.
\end{align}   
Here, to simplify the calculation,  we can choose the magnetic field along the $z$ direction. The direction of the electric field and the temperature gradient are perpendicular to the $z$ axis {\it, i.e.} it is in the $x-y$ plane. Under these conditions, the components of the electric current in the $x-y$ plane are given as,
\begin{widetext}
\begin{align}
 j_x = &  \sum_{i} \frac{q_i g_i}{3} \int \frac{d^3|\vk_i|}{(2\pi)^3} v_i^2 \frac{\tau_R^i}{(1+ \chi_i + \chi_i^2)(1+\chi_i)}
\Big[-q_i\Big\{ \Big(
(1+\chi_i)
+\frac{\chi_i(1+ \chi_i - \chi_i^2)}{(1+ \chi_i^2)}\Big)E^x 
\pm \Big(\chi_i(1+\chi_i)+
\frac{\chi_i^2(2+ \chi_i)}{(1+ \chi_i^2)} \Big)
E^y\Big\}\nn\\
&+ \frac{(\omega_i-{\rm b}_i \xi)}{T}\Big\{(1+\chi_i)
\frac{\del {T}}{\del x}
\pm\chi_i(1+\chi_i)
\frac{\del {T}}{\del y}\Big\}\Big]\frac{\partial f^0_i}{\partial \om_i},
 \label{equnew38}
\end{align}
and,
\begin{align}
 j_y = &  \sum_{i} \frac{q_i g_i}{3} \int \frac{d^3|\vk_i|}{(2\pi)^3} v_i^2 \frac{\tau_R^i}{(1+ \chi_i + \chi_i^2)(1+\chi_i)}
\Big[-q_i\Big\{ \Big(
(1+\chi_i)
+\frac{\chi_i(1+ \chi_i - \chi_i^2)}{(1+ \chi_i^2)}\Big)E^y 
\pm \Big(\chi_i(1+\chi_i)-
\frac{\chi_i^2(2+ \chi_i)}{(1+ \chi_i^2)} \Big)
E^x\Big\}\nn\\
&+ \frac{(\omega_i-{\rm b}_i \xi)}{T}\Big\{(1+\chi_i)
\frac{\del {T}}{\del y}
\mp\chi_i(1+\chi_i)
\frac{\del {T}}{\del x}\Big\}\Big]\frac{\partial f^0_i}{\partial \om_i}.
 \label{equnew39}
\end{align}
Eq.\eqref{equnew38} and \eqref{equnew39} can be written in a compact form by introducing the following integrals,
\begin{align}
&H_{1_i} = \frac{g_i }{3T} \int \frac{d^3|\vk_i|}{(2\pi)^3}
\frac{\vk^2_i}{\om_i^2} f^0_i(1-f^0_i) \tau_R^i
\times \frac{(1+ \chi_i^2)+\chi_i(2+ \chi_i)}{(1+\chi_i) \left(1+\chi_i^2\right) \left(1+ \chi_i + \chi_i^2\right)}~,\label{equnew40} \\
&H_{2_i} = \frac{g_i}{3T}  \int \frac{d^3|\vk_i|}{(2\pi)^3}
\frac{\vk^2_i}{\om_i^2} f^0_i(1-f^0_i) \tau_R^i
\times \chi_i\frac{(1+\chi_i)(1+ \chi_i^2)+\chi_i(2+ \chi_i)}{(1+\chi_i) \left(1+\chi_i^2\right) \left(1+ \chi_i + \chi_i^2\right)}~,\label{equnew41}\\
&H_{3_i} = \frac{g_i}{3T}   \int \frac{d^3|\vk_i|}{(2\pi)^3}\frac{\vec{k}^2_i}{\om_i^2}(\omega_i-{\rm b}_i \xi)f^0_i(1 - f^0_i) \tau_R^i 
\times \frac{1}{(1+\chi_i + \chi_i^2)}~,\label{equnew42}\\    
&H_{4_i} = \frac{g_i}{3T} \int \frac{d^3|\vk_i|}{(2\pi)^3}\frac{\vec{k}^2_i}{\om_i^2}(\omega_i-{\rm b}_i \xi)f^0_i(1 - f^0_i) \tau_R^i
\times \frac{\chi_i}{(1+\chi_i + \chi_i^2)}\label{equnew43}~.
 \end{align}
 Here, it is important to note that the expressions are obtained in the limit of a slowly varying magnetic field, for which we approximated the decay parameter as the inverse of cyclotron frequency, \i.e., $\tau_B = \frac{\om_i}{q_iB}$. Furthermore, $H_{2_i}$ and $H_{4_i}$ should have explicit sign dependency from the electric charge of the particle due to $\chi_i$ in the numerator. However, this information vanishes due to the approximation. Therefore, we use the minus (plus) sign in $H_{2_i}$ and $H_{4_i}$ for negatively (positively) charged particles and antiparticles for the numerical estimations.
The integrals as given in Eq.\eqref{equnew40}-Eq.\eqref{equnew43} allows us to write  Eq.\eqref{equnew38} and  Eq.\eqref{equnew39}, respectively, as 
\begin{align}
 j_x = \sum_i q_i^2H_{1_i}E_x+\sum_i q_i^2H_{2_i}E_y -\frac{1}{T}\sum_{a}q_iH_{3_i}\frac{d T}{dx}-\frac{1}{T}\sum_{a}q_iH_{4_i}\frac{d T}{dy},
 \label{equnew44}
\end{align}
and,
\begin{align}
 j_y = \sum_i q_i^2H_{1_i}E_y-\sum_i q_i^2H_{2_i}E_x -\frac{1}{T}\sum_{a}q_iH_{3_i}\frac{d T}{dy}+\frac{1}{T}\sum_{a}q_iH_{4_i}\frac{d T}{dx}.
 \label{equnew45}
\end{align}
Here, in the presence of a magnetic field, the magneto-Seebeck coefficient ($S_B$) can be determined by setting $j_x = j_y=0$ so that the electric field becomes proportional to the temperature gradient. For $j_x=0$ and $j_y=0$, we can solve Eq.\eqref{equnew44} and \eqref{equnew45} to get $E_x$ and $E_y$ in terms of temperature gradients $\frac{dT}{dx}$ and  $\frac{dT}{dy}$ as,
\begin{align}
 E_x &  = \frac{\sum_i q_i^2H_{1_i}\sum_iq_iH_{3_i}+\sum_i q_i^2H_{2_i}\sum_iq_iH_{4_i}}{T\bigg[\bigg(\sum_i q_i^2H_{1_i}\bigg)^2+\bigg(\sum_i q_i^2H_{2_i}\bigg)^2\bigg]}\frac{dT}{dx}
 +\frac{\sum_i q_i^2H_{1_i}\sum_iq_iH_{4_i}-\sum_i q_i^2H_{2_i}\sum_iq_iH_{3_i}}{T\bigg[\bigg(\sum_i q_i^2H_{1_i}\bigg)^2+\bigg(\sum_i q_i^2H_{2_i}\bigg)^2\bigg]}\frac{dT}{dy},
 \label{equnew46}
\end{align}
and,
\begin{align}
 E_y &  = \frac{\sum_i q_i^2H_{2_i}\sum_iq_iH_{3_i}-\sum_i q_i^2H_{1_i}\sum_iq_iH_{4_i}}{T\bigg[\bigg(\sum_i q_i^2H_{1_i}\bigg)^2+\bigg(\sum_i q_i^2H_{2_i}\bigg)^2\bigg]}\frac{dT}{dx}
 +\frac{\sum_i q_i^2H_{1_i}\sum_iq_iH_{3_i}+\sum_i q_i^2H_{2_i}\sum_iq_iH_{4_i}}{T\bigg[\bigg(\sum_i q_i^2H_{1_i}\bigg)^2+\bigg(\sum_i q_i^2H_{2_i}\bigg)^2\bigg]}\frac{dT}{dy}.
 \label{equnew47}
\end{align}
\end{widetext}
Eq.
\eqref{equnew46} and \eqref{equnew47} can be written in a compact form in the following way,
\begin{align}
 \begin{pmatrix}
E_x \\
\\
E_y 
\end{pmatrix}= \begin{pmatrix}
S_B & NB \\
\\
-NB & S_B 
\end{pmatrix}\begin{pmatrix}
\frac{dT}{dx} \\
\\
\frac{dT}{dy} 
\end{pmatrix},
\end{align}
Here, one can identify the magneto-Seebeck coefficient as, 

\begin{align}
 S_B & = \frac{\sum_i q_i^2H_{1_i}\sum_iq_iH_{3_i}+\sum_i q_i^2H_{2_i}\sum_iq_iH_{4_i}}{T\bigg[\bigg(\sum_i q_i^2H_{1_i}\bigg)^2+\bigg(\sum_i q_i^2H_{2_i}\bigg)^2\bigg]}\nonumber\\
 & = \frac{(\sigma_{el}/T)(\mathcal{I}_{31}/T^2)+(\sigma_{H}/T)(\mathcal{I}_{42}/T^2)}{(\sigma_{el}/T)^2+(\sigma_{H}/T)^2},
 \label{equnew49}
\end{align}
and the normalized Nernst coefficient ($NB$) is given as,
\begin{align}
 NB & = \frac{\sum_i q_i^2H_{1_i}\sum_iq_iH_{4_i}-\sum_i q_i^2H_{2_i}\sum_iq_iH_{3_i}}{T\bigg[\bigg(\sum_i q_i^2H_{1_i}\bigg)^2+\bigg(\sum_i q_i^2H_{2_i}\bigg)^2\bigg]}\nonumber\\
 & = \frac{(\sigma_{el}/T)(\mathcal{I}_{42}/T^2)-(\sigma_{H}/T)(\mathcal{I}_{31}/T^2)}{(\sigma_{el}/T)^2+(\sigma_{H}/T)^2}.
 \label{equnew50}
 \end{align}

Now, we have identified the Ohmic-like component of electrical conductivity
in the presence of a magnetic field and the Hall-like component of electrical conductivity as $\sigma_{el}=\sum_iq_i^2H_{1_i}$ and $\sigma_{H}=\sum_iq_i^2H_{2_i}$ respectively \cite{Singh:2024emy}. The integrals $\mathcal{I}_{31}$ and $\mathcal{I}_{42}$ in Eqs.\eqref{equnew49} and \eqref{equnew50} are defined as  $\mathcal{I}_{31} = \sum_iq_iH_{3_i}$ and $\mathcal{I}_{42}\equiv \sum_iq_iH_{4_i}$.
Note that in the absence of a magnetic field, integrals $H_{2_i}$ and $H_{4_i}$ turn out to be zero. Hence, the normalized Nernst coefficient vanishes in the absence of a magnetic field, and the magneto-Seebeck coefficient turns into the Seebeck coefficient in the absence of a magnetic field as given in Eq.\eqref{equnew11}. 
Finally, we can calculate magneto-Thomson coefficient ($Th_B$) from the magneto-Seebeck coefficient $S_B$ as
\begin{align}\label{Mag-Thomson}
    Th_B = T\frac{dS_B}{dT}.
\end{align}
In the absence of a magnetic field, the coefficient $Th_B$ reduces to the coefficient $Th$. The transverse Thomson coefficient ($Th_N$) also originates in the medium because of the presence of the finite value of the normalized Nernst coefficient. Phenomenologically, $Th_N$ is expected to occur when a charge current, temperature gradient, and magnetic field are oriented orthogonally to each other in any conducting medium. Unlike the $Th_B$, the coefficient $Th_N$ not only depends on the temperature derivative of the leading thermoelectric coefficient (for this case $NB$) but also on its magnitude as~\cite{takahagi}
\begin{align}\label{Transverse-Thomson}
    Th_N = T\frac{d(NB)}{dT} + 2NB.
\end{align}
In the absence of a magnetic field, the coefficient $Th_N$ vanishes because of the vanishing $NB$. The first term in the above equation is the dynamic part, which tells how the magneto-transport properties of the medium evolve as it cools or heats. The second term tells how strong the transverse thermoelectric response is due to the magnetic field at the current temperature, regardless of how fast it is changing.
\subsection{Estimation of relaxation time for QGP medium}
In our current work, we use a quasiparticle model proposed by Gorenstein and Yang~\cite{Gorenstein:1995vm} to estimate the QGP equation of state numerically. In this phenomenological model, lattice QCD results are reproduced by attributing effective thermal masses to the partons, where the thermal mass $m(T)$ arises from the interactions among the partons. The thermodynamic consistency is achieved by introducing a bag constant arising from vacuum energy~\cite{Bannur:2006hp}. The thermal mass of the ith flavor of quarks is~\cite{Srivastava:2010xa}
\begin{align}
    m_{iT}^2(T,\mu_B) = \frac{N_c^2 - 1}{8N_c}\Big( T^2 + \frac{\mu_B^2}{9 \pi^2}\Big)g^2(T,\mu_B).
\end{align}
The effective mass of the gluon $(m_g)$ in this model can be represented as
\bea
m_{g}^2(T,\mu_B) = \frac{N_c}{6} g^2(T,\mu_B) T^2 \left(1+\frac{N_f + \frac{\mu_B^2}{\pi^2 T^2}}{6}\right).
\eea
$N_c, N_f$ represent the number of color degrees of freedom and the number of quark flavors, respectively. Here, $g^2(T,\mu_B) = 4\pi \alpha_s(T,\mu_B)$, we consider $\alpha_s(T,\mu_B)$ as a two-loop running coupling constant~\cite{Srivastava:2010xa}.
The dispersion relation of the particle having energy $\om_i$ and momentum $k_i$ is $\om_i^{2}(k_i, T) = k_i^{2} + m_i^{2}(T)$. Where $m_i$ is the total effective mass of the ith quark flavor and parameterized as $m_i^2 = m_{i0}^2 + \sqrt{2}m_{i0}m_{iT} + m_{iT}^2$.
$m_{i0}$ and $m_{iT}$ are the bare masses. In the presence of a magnetic field, the $\alpha_s(T,\mu_B)$ modifies as~\cite{Ayala:2018wux, Sahoo:2024yud}
\begin{align}
    \alpha_s(T,\mu_B,eB) = \frac{\alpha_s(T,\mu_B)}{1 + \gamma~\alpha_s(T,\mu_B) \ln \left ( \frac{\frac{1}{\Lambda_T^2}(T^2+\frac{\mu_B^2}{9 \pi^2})}{\frac{1}{\Lambda_T^2}(T^2+\frac{\mu_B^2}{9 \pi^2})+ |eB|} \right)}.
\end{align}
Here, $\gamma = \frac{11N_c - 2N_f}{12\pi}$ and $\Lambda_T$ is the QCD scale-fixing parameter taken to be 0.115 GeV in our calculations. For the case of $eB$ = 0, $\alpha_s(T,\mu_B,eB)$ reduces to $\alpha_s(T,\mu_B)$. 
Now, for the relaxation time of quarks, we use a momentum-independent expression obtained for QCD matter~\cite{Hosoya:1983xm}
\begin{align}
    \tau_R = \frac{1}{5.1T\alpha_s^{2}\log(1/\alpha_s)[1+0.12(2N_f+1)]}.
\end{align}
For numerical estimation, the value of the strong coupling constant is taken to be fixed at $\alpha_s = 0.5$. 
\begin{figure*}
	\centering
	\includegraphics[scale=0.40]{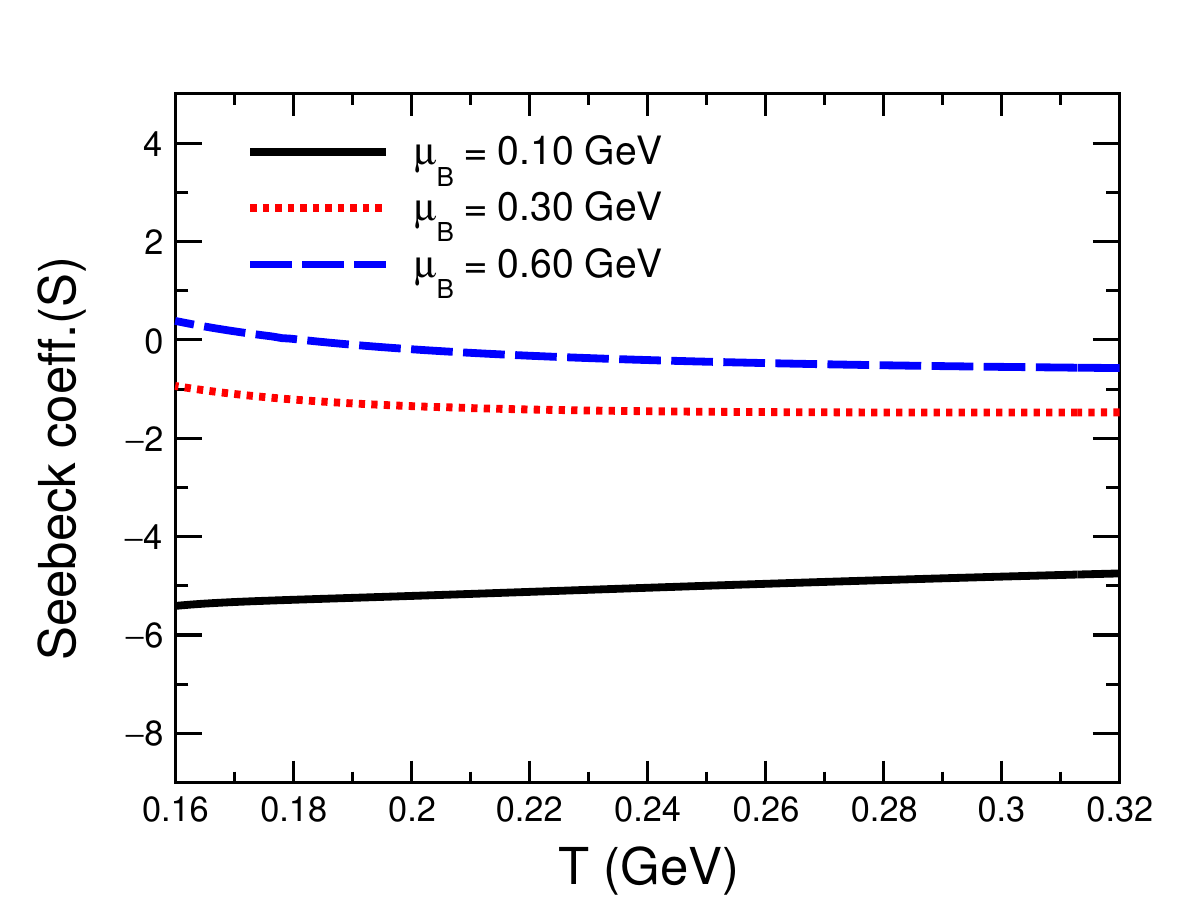}
        \includegraphics[scale=0.40]{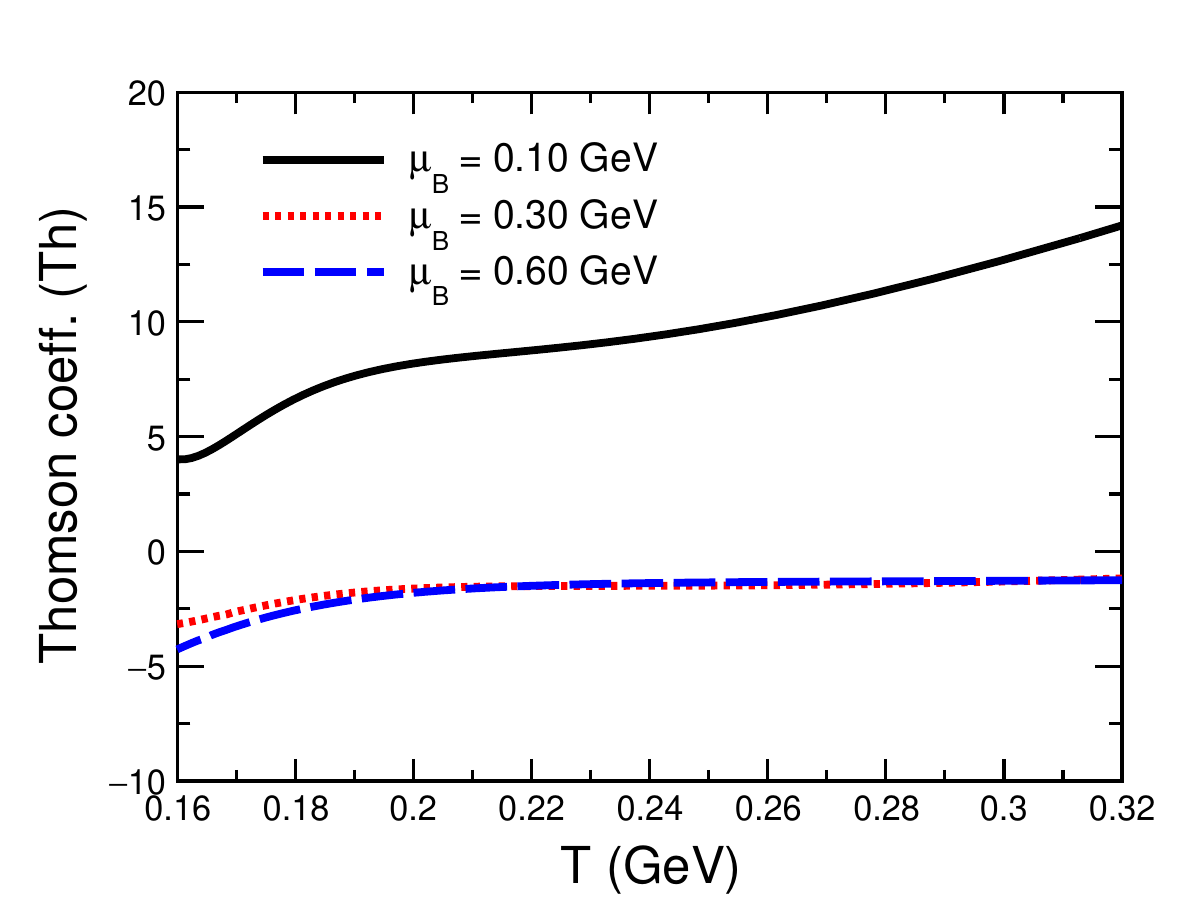}
\caption{Left: Seebeck coefficient ($S$), right: Thomson coefficient ($Th$) as a function of temperature ($T$) for different values of $\mu_B$ = 0.10, 0.30, and 0.60 GeV.}
	\label{Seebeck}
\end{figure*}

\section{Results and Discussion}\label{Sec-results}
We have calculated the thermoelectric coefficients for the QGP medium for both the cases of zero and non-zero external magnetic fields using the QPM model. The left panel of Fig.~\ref{Seebeck} represents the Seebeck coefficient $S$ as a function of temperature $T$ for different values of baryon-chemical potential $\mu_B$ = 0.10, 0.30, 0.60 GeV. The solid black line represents the case of $\mu_B$ = 0.10 GeV. On the other hand, the red dotted line and the blue dashed line represent the case for $\mu_B$ = 0.30, 0.60 GeV, respectively. In our calculations, the sign of $S$ is dependent on the numerator of Eq.~(\ref{equnew11}). The value of enthalpy per particle $\xi$ plays a crucial role in the flow of heat current $\mathcal{I}_{1}$. The gluons do not directly contribute to $\mathcal{I}_{1}$ due to their zero electric charge but enter through their contribution to the enthalpy of the medium. It is to be noted that $\xi$ increases as the temperature $T$ of the medium increases, but it decreases with the increase of the $\mu_B$. This means that for the fixed value of $T$, $\xi$ is higher for the lower value of $\mu_B$. Therefore, for lower values of $\mu_B$ (= 0.10, 0.30 GeV), the $S$ is negative throughout the temperature range, which is due to a larger value of enthalpy per particle $\xi$ as compared to the single-particle energy $\omega_i$; hence the term $\omega_i-{\rm b}_i \xi$ becomes negative in Eq.~(\ref{equnew11}). On the other hand, for the higher value of $\mu_B$ (= 0.60 GeV), the $S$ is slightly positive at lower values of $T$; later on, it turns negative. For a particular medium, the sign of $S$ describes the alignment of the induced electric field with respect to the temperature gradients present in the medium. Here, the temperature dependence of $S$ gives rise to the non-zero Thomson coefficient in the medium. The right panel of Fig.~\ref{Seebeck} represents the Thomson coefficient $Th$ as a function of temperature $T$ for different values of baryon-chemical potential. The solid black line represents the case of $\mu_B$ = 0.10 GeV. On the other hand, the red dotted line and the blue dashed line represent the case for $\mu_B$ = 0.30, 0.60 GeV, respectively. The Thomson coefficient describes the continuous
absorption or release of heat in the charge-carrying medium in the presence of temperature gradients. In condensed matter systems, for a material having positive $Th$, the heat is absorbed when current flows from a region of lower temperature to a higher temperature, such as copper, silver, zinc, etc. On the contrary, the heat is released if $Th$ is negative, such as cobalt, nickel, bismuth, etc.~\cite{book}. Here, we observe that for a QGP medium having a low value of $\mu_B$ (= 0.10 GeV), the $Th$ is positive and increases as $T$ increases. Whereas for other values of $\mu_B$ (= 0.30, 0.60 GeV), $Th$ is negative for the whole temperature range. It slightly increases with $T$ initially, then gets saturated at higher values of $T$. This behavior of $Th$ is not independent of $S$ as clearly mentioned in Eq.~(\ref{Thomson}). In a trivial sense, we can estimate that the higher the slope of $S$ in the left panel, the higher the value of $Th$ in the right panel of Fig.~\ref{Seebeck}. The non-zero values of $Th$ may affect the cooling of the medium during the evolution. 
\begin{figure*}
	\centering
	\includegraphics[scale=0.40]{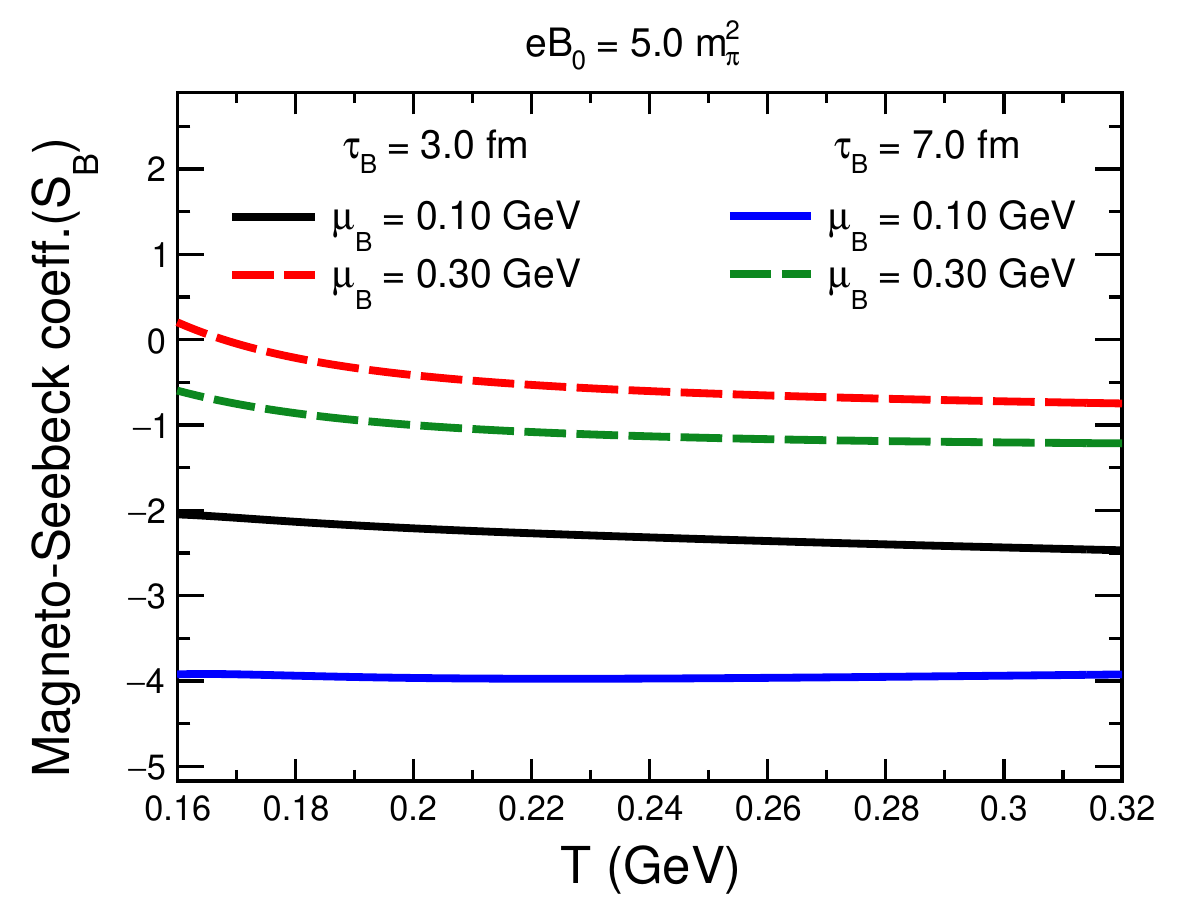}
        \includegraphics[scale=0.40]{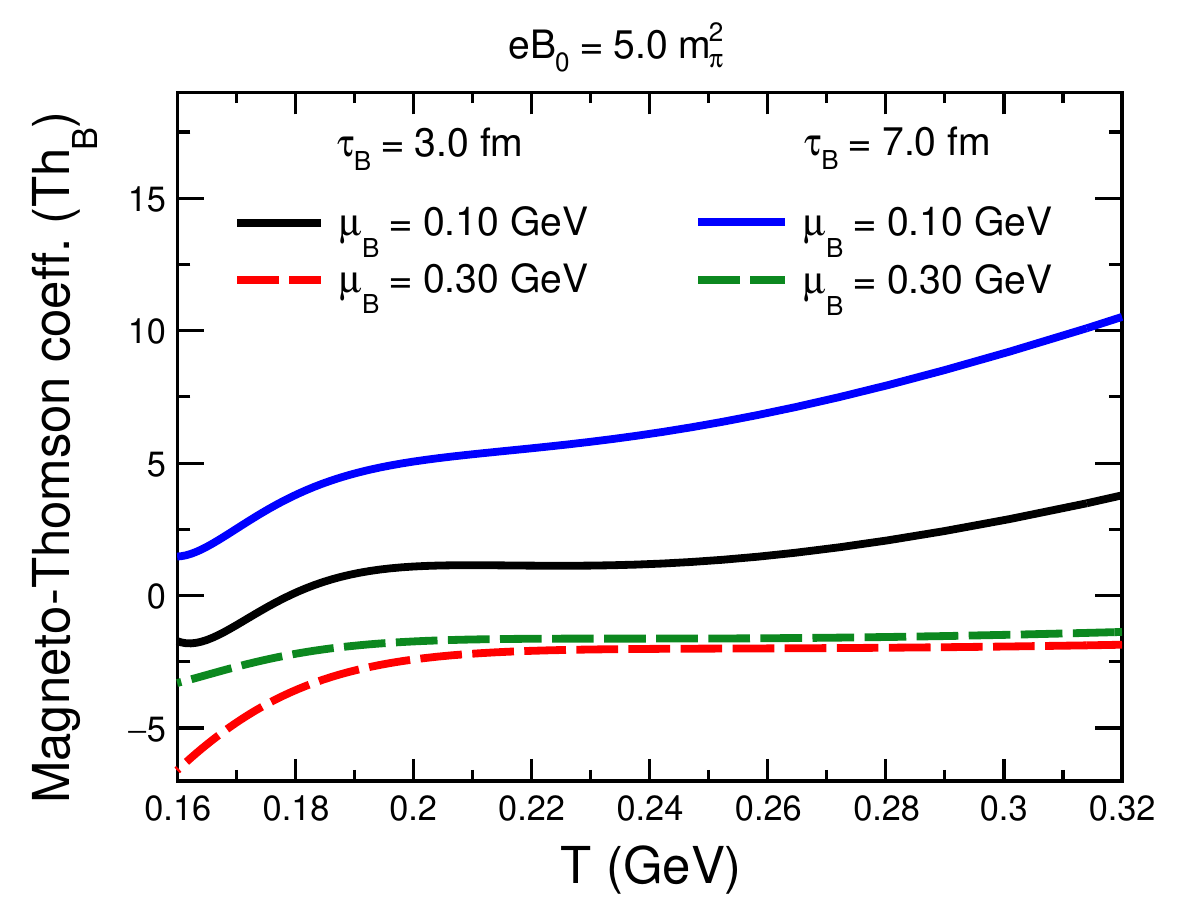}
    \caption{Left: magneto-Seebeck coefficient ($S_B$), right: magneto-Thomson coefficient ($Th_B$) as a function of temperature ($T$) for different values of $\mu_B$ = 0.10 and 0.30 GeV and at magnetic field $(eB_0) = 5.0~m_\pi^2$ with decay parameter $(\tau_B)$ = 3.0, 7.0 fm.}
\label{Magnetoseebeck}
\end{figure*}
\begin{figure*}
	\centering
	\includegraphics[scale=0.40]{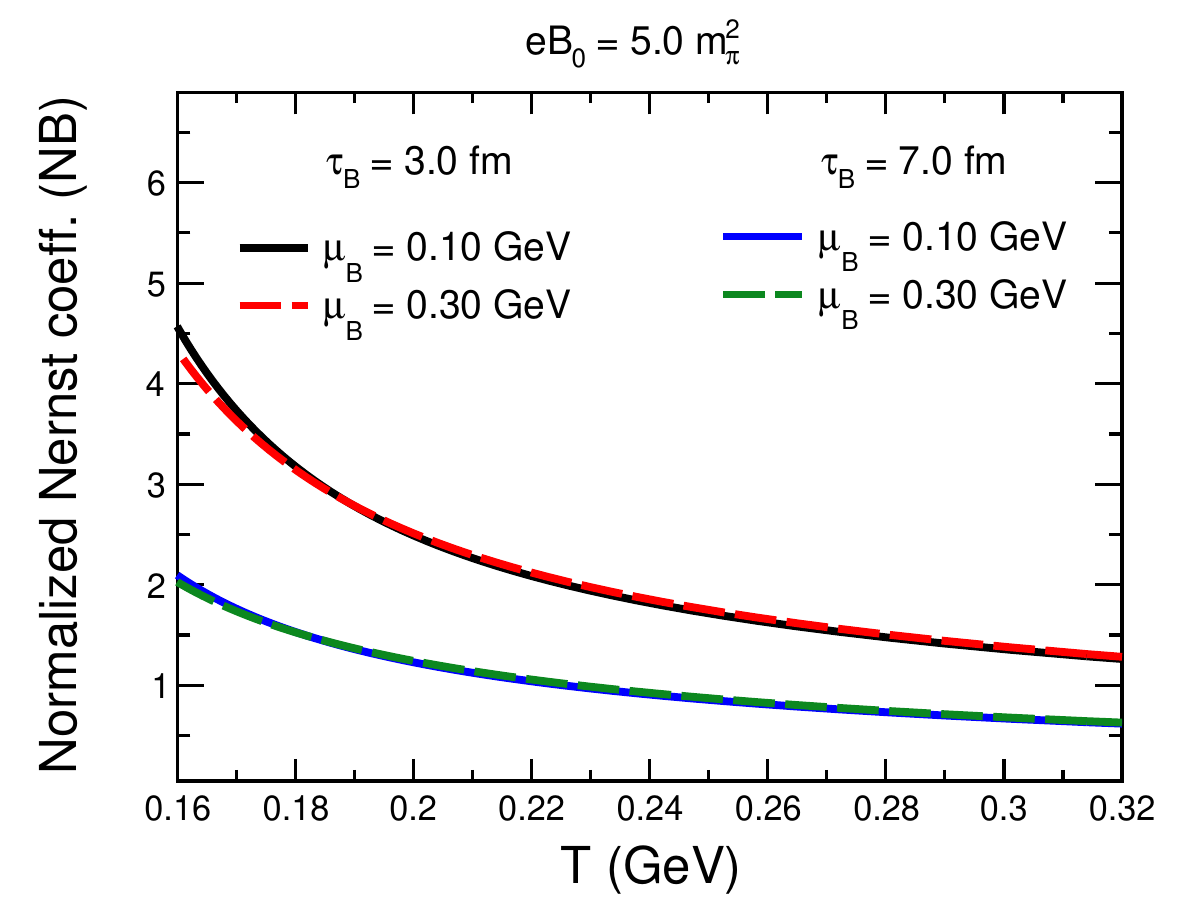}
        \includegraphics[scale=0.40]{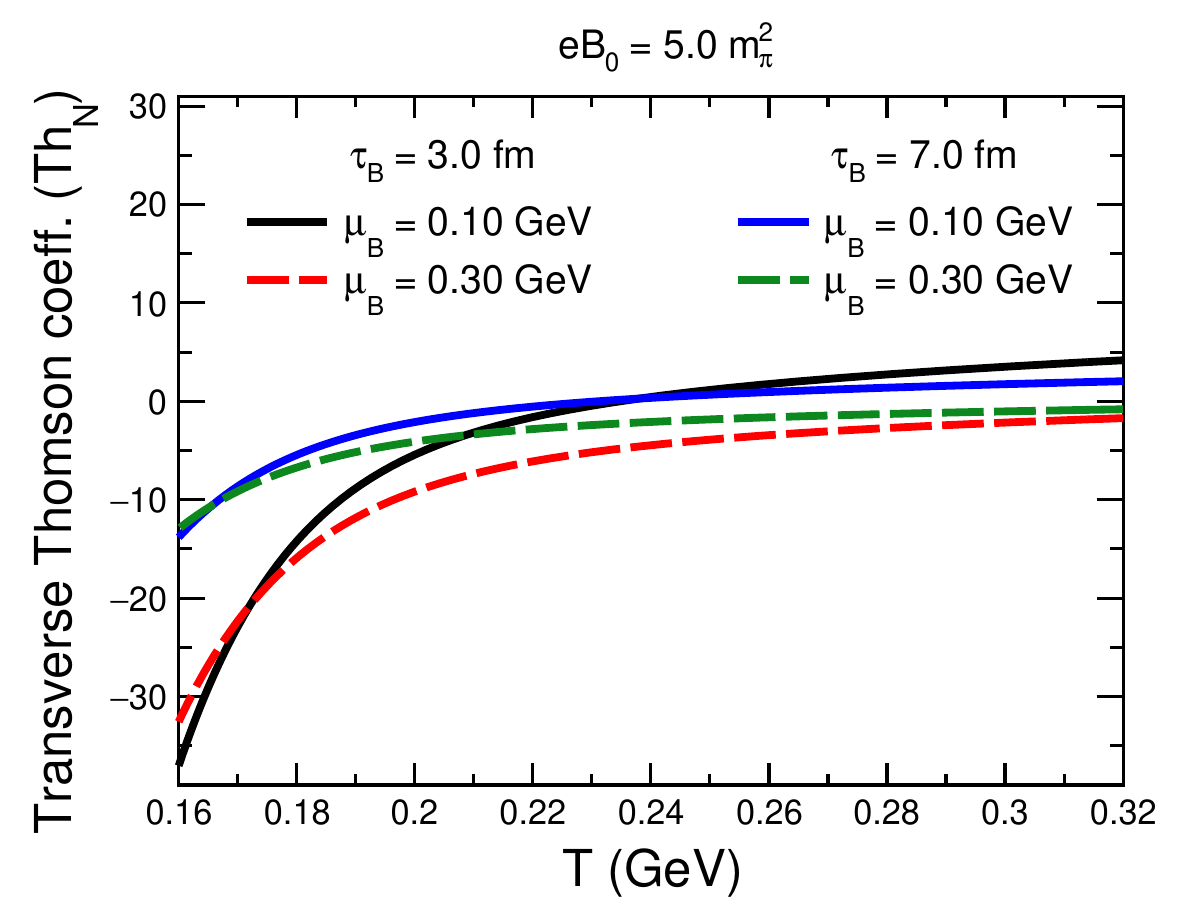}
    \caption{Left: normalized Nernst coefficient ($NB$), right: transverse Thomson coefficient ($Th_N$) as a function of temperature ($T$) for different values of $\mu_B$ = 0.10 and 0.30 GeV and at magnetic field $(eB_0) = 5.0~m_\pi^2$ with decay parameter $(\tau_B)$ = 3.0, 7.0 fm.}
\label{Nernst}
\end{figure*}
In Fig.~\ref{Magnetoseebeck}, we represent the results for the thermoelectric coefficient of the medium for the case of nonzero magnetic fields. To account for the simultaneous evolution of the magnetic field and temperature, we adopt a mapping between proper time and temperature based on an ideal magnetohydrodynamic (MHD) framework, as discussed in our earlier work~\cite{Singh:2024emy}. In that study, the temperature and magnetic field were both expressed as functions of proper time for different decay parameters. For the present analysis, which focuses on higher-order thermoelectric coefficients in a static QGP scenario, we extract consistent pairs of magnetic field and temperature values corresponding to the same proper time using these decay profiles. This allows us to plot thermoelectric coefficients as a function of temperature while implicitly incorporating the effect of time-varying magnetic fields for different decay rates.
The left panel represents the magneto-Seebeck coefficient $S_B$ as a function of temperature $T$ for two different values of baryon-chemical potential $\mu_B$ = 0.10, 0.30 GeV. The initial value of the magnetic field ($eB_0$) is taken to be 5 $m_\pi^2$ along with two different magnetic decay parameters $\tau_B$ at 3, 7 fm. The solid black and blue lines represent the case of $\mu_B$ = 0.10 GeV for $\tau_B$ = 3, 7 fm, respectively. On the other hand, the red dashed and green dashed lines represent the case for $\mu_B$ = 0.30 for $\tau_B$ = 3, 7 fm, respectively. Unlike the case of zero magnetic fields, we can see in Eq.~(\ref{equnew49}) that, along with the Ohmic-like components of electrical conductivity and heat current, the Hall-like components also contribute. In the QGP medium, only quarks contribute to the Hall-like components as they experience the Lorentz force in the magnetic field. To study the Seebeck effect in the presence of a magnetic field, we now discuss the case of $\mu_B = 0.10$ GeV. For this value of $\mu_B$, both components of electrical conductivity ($\sigma_{el}, \sigma_H$) contribute to the magneto-Seebeck coefficient, which is positive throughout the full temperature range. On the other hand, the Ohmic-like component of heat current ($\mathcal{I}_{31}$) is negative, but the Hall-like component ($\mathcal{I}_{42}$) is positive because of the multiplied factor $\chi_i$. Hence, the product $(\sigma_{el}/T)(\mathcal{I}_{31}/T^2)$ is negative, but $(\sigma_{H}/T)(\mathcal{I}_{42}/T^2)$ is positive, and this results into decrease in value of $S_B$. In the absence of a magnetic field, all the Hall-like components vanish; hence, $S_B$ turns the same as $S$. In the left panel of Fig.~\ref{Magnetoseebeck}, we can see that the different magnetic decay parameters $\tau_B$ have different effects on $S_B$. For the case of a fast decaying magnetic field ($\tau_B = 3$ fm),  $S_B$ reduces more as compared to the case of a slowly decaying magnetic field ($\tau_B = 7$ fm). This is because of the multiplied factor $\chi_i$, which is higher in magnitude for the lower value of $\tau_B$. Now, the right panel of Fig.~\ref{Magnetoseebeck} represents the magneto-Thomson coefficient $Th_B$ as a function of temperature $T$ for two different values of baryon-chemical potential $\mu_B$ = 0.10, 0.30 GeV. As we mentioned above, the discussion for $S_B$, $Th_B$ also decreases in the presence of a magnetic field. From Eq.~(\ref{Mag-Thomson}), we can see that the higher the slope of $S_B$ in the left panel, the larger the value of $Th_B$. The complete saturation of $S_B$ over the temperature leads to the vanishing $Th_B$. The similar effect of $\tau_B$ is reflected in $Th_B$ as it is for $S_B$. For a fast decaying magnetic field, the value of $Th_B$ is lower than compared of a slowly decaying magnetic field. 
\begin{figure*}
	\centering
	\includegraphics[scale=0.40]{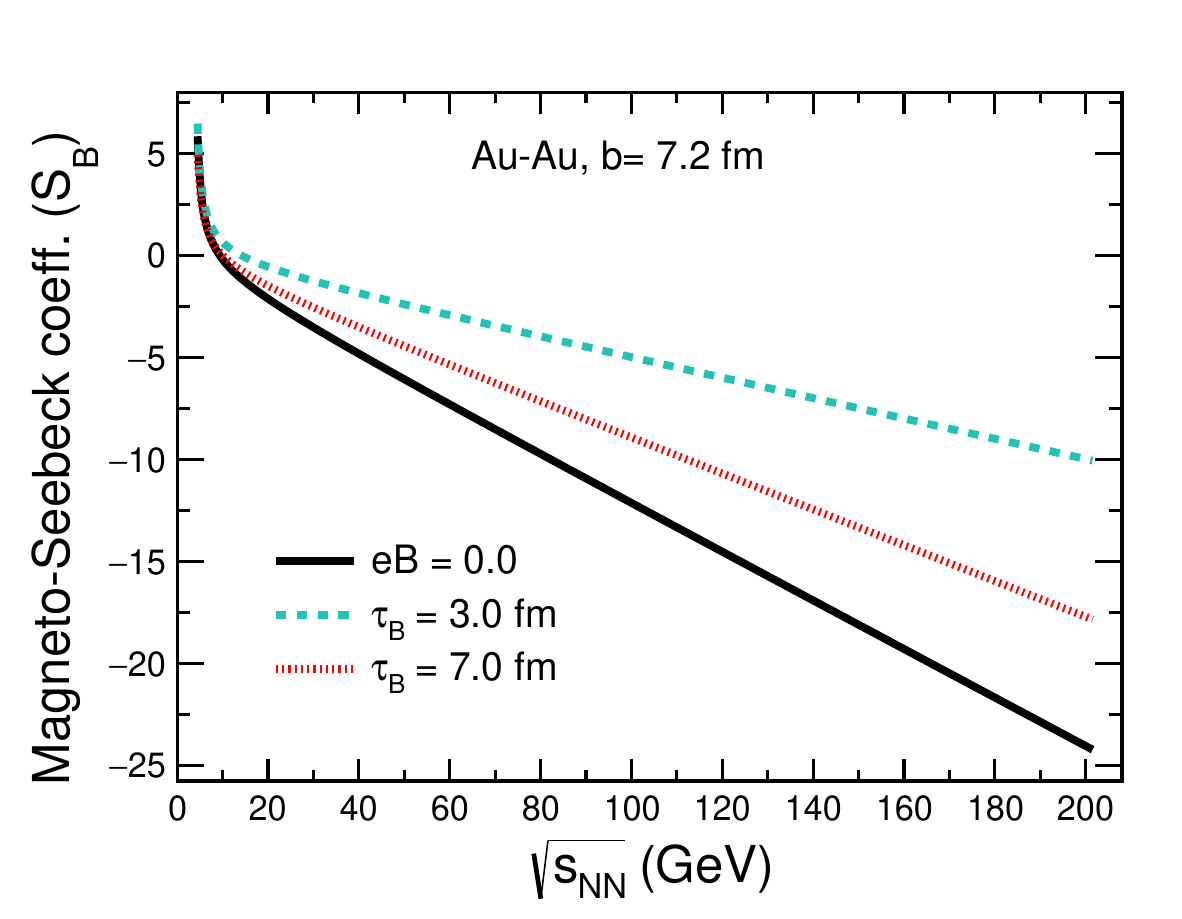}
        \includegraphics[scale=0.40]{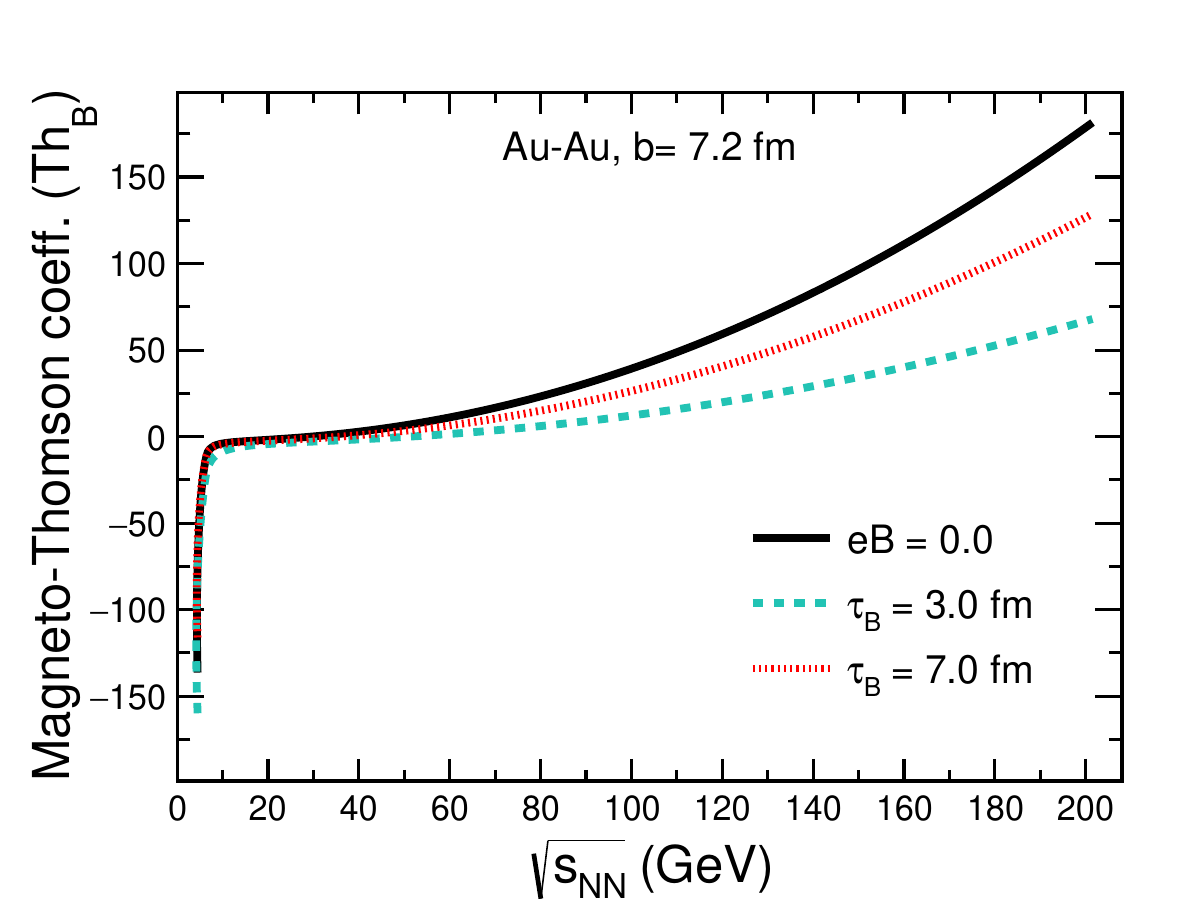}
    \caption{Left: magneto-Seebeck coefficient ($S_B$), right: magneto-Thomson coefficient ($Th_B$) as a function of center of mass energy ($\sqrt{s_{NN}}$) for vanishing magnetic field $(eB) = 0.0$ and different values of decay parameter $(\tau_B)$ = 3.0, 7.0 fm.}
\label{E0}
\end{figure*}
\begin{figure*}
	\centering
	\includegraphics[scale=0.40]{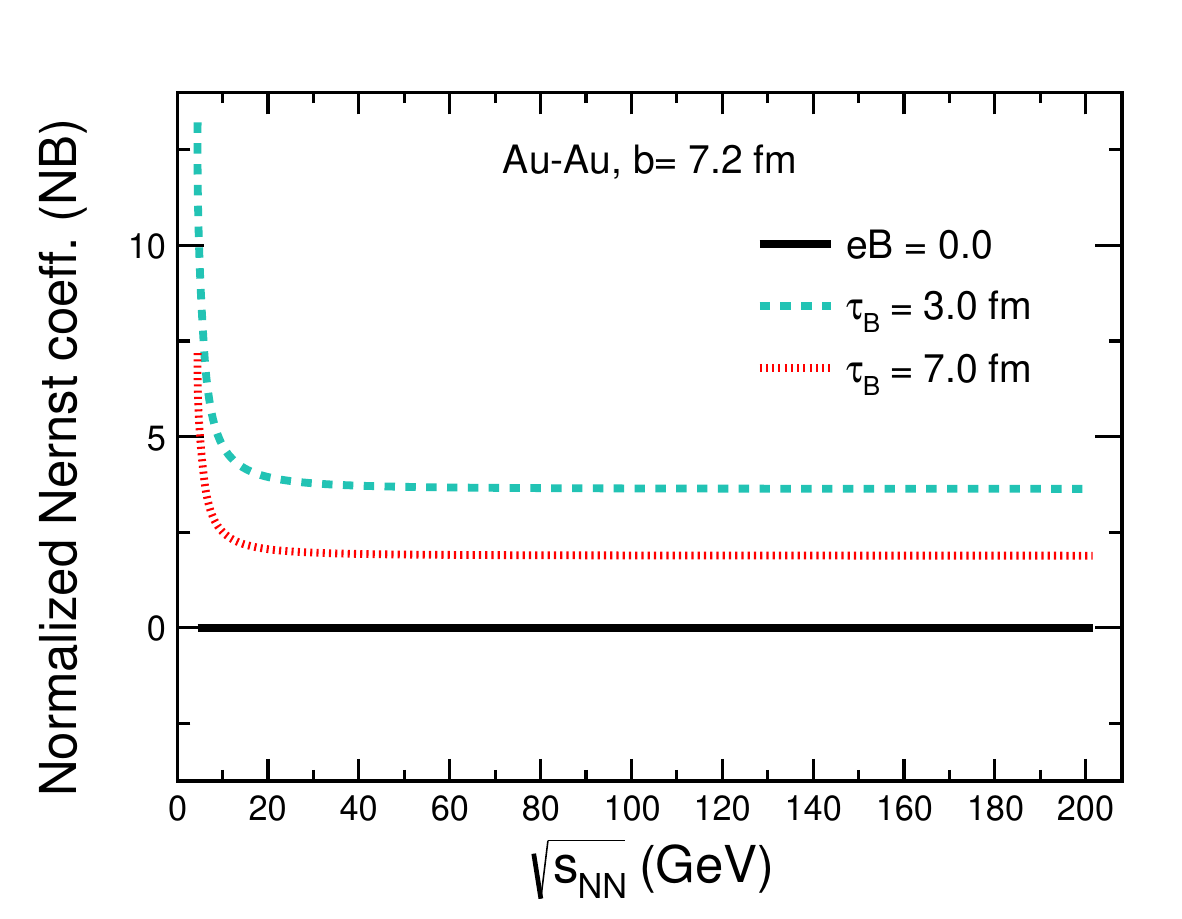}
        \includegraphics[scale=0.40]{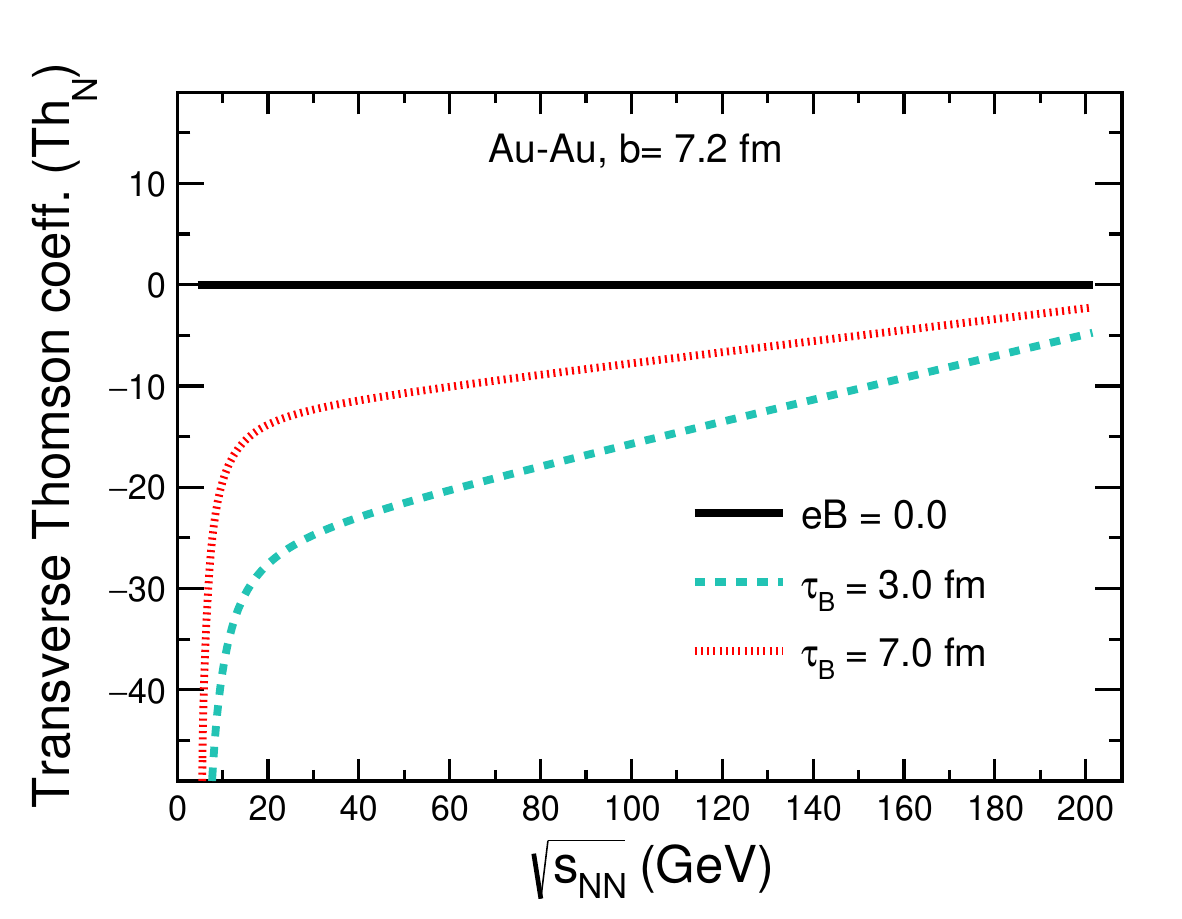}
    \caption{Left: normalized Nernst coefficient ($NB$), right: transverse Thomson coefficient ($Th_N$) as a function of center of mass energy ($\sqrt{s_{NN}}$) for vanishing magnetic field $(eB) = 0.0$ and different values of decay parameter $(\tau_B)$ = 3.0, 7.0 fm.}
\label{Transverse}
\end{figure*}
The left panel of Fig.~\ref{Nernst} represents the normalized Nernst coefficient $NB$ as a function of temperature for two different values of $\mu_B$ = 0.10, 0.30 GeV and $\tau_B$ = 3, 7 fm. We observe that $NB$ decreases as the temperature increases for all values of $\mu_B$ and $\tau_B$. Here, both $\sigma_{el}, \sigma_H$ contribute to the $NB$ are positive throughout the temperature range discussed here. On the other hand, $\mathcal{I}_{31}$ is negative, but the $\mathcal{I}_{42}$ is positive because of the multiplication factor $\chi_i$. Hence, the product $(\sigma_{H}/T)(\mathcal{I}_{31}/T^2)$ is positive, along with the product $(\sigma_{H}/T)(\mathcal{I}_{42}/T^2)$ which is also positive, and this results into positive values of $NB$. The values of $NB$ are higher for the faster decaying magnetic field. This is because of the fact that $\chi_i$ is higher for the smaller value of $\tau_B$. In the absence of a magnetic field, all the Hall-like component vanishes; hence, $NB$ also vanishes. The right panel of Fig.~\ref{Nernst} represents the transverse Thomson coefficient $NB$ as a function of temperature for two different values of $\mu_B$ and $\tau_B$. The results demonstrate that the coefficient $Th_N$ is highly sensitive to both $\mu_B$ and $\tau_B$, with more negative values observed at lower temperatures. For the higher values of temperatures, the coefficient $Th_N$ seems to be saturated and approaches zero. In a trivial sense, this can be estimated from the coefficient $NB$ that has the minimum slope towards the higher temperature range and the smaller values in magnitude. Therefore, for the low temperature region, the derivative term in Eq.~\ref{Transverse-Thomson} dominates over $2NB$, whereas at high temperatures, where $NB$ seems to saturate, the term $2NB$ dominates. 

To express $Th_B$ and $Th_N$ as a function of center of mass energy $\sqrt{s_{NN}}$ (in GeV) we follow the parameterization~\cite{Cleymans:2005xv}
\begin{align}\label{par}
    T(\mu_{B}) &= 0.166 - 0.139 \mu_B^2 - 0.053 \mu_B^4, \nn\\
    \mu_{B} &= \frac{1.308}{1 + 0.273\sqrt{s_{NN}}}.
\end{align}
Also, the magnetic field is parameterized as~\cite{Tuchin:2013ie}, 
\begin{align}
    B = \frac{\sqrt{s_{NN}}}{8\pi ~m_N}Ze\frac{b}{R^3} \exp{\left(-\frac{\tau}{\tau_B}\right)}.
\end{align}
Here, $R$ is the radius of colliding ions with electric charge $Ze$, $b$ is the impact parameter, and $m_N = 0.938$~GeV is the nucleon mass. For Au-Au collision at $\sqrt{s_{NN}} = 200$~GeV, with $R=6.38$~fm, $b=7.2$~fm, and $Z = 79$ we get $eB_0 \approx 3 ~m_\pi^2$ at $\tau = 0$, which is in accordance with the results for RHIC energies~\cite{Deng:2012pc}. The left panel of Fig.~\ref{E0} represents the $S_B$ as a function of $\sqrt{s_{NN}}$. The black solid line represents the case of zero magnetic fields, whereas the cyan dashed line and red dotted line represent the case of the non-zero magnetic field for $\tau_B$ = 3, 7 fm, respectively. We observe that for lower values of center of mass energies, nearly up to 10 GeV, the magnetic field does not show any significant effect on $S_B$. Later on, as the values of $\sqrt{s_{NN}}$ increase, the effect of the magnetic field, along with different decay parameters, also increases. The faster the decay of the magnetic field, the lower the value of $S_B$. Here, we also observe that the $S_B$ is positive at lower energies; it vanishes nearly at $\sqrt{s_{NN}} \approx 20$~GeV and then starts increasing towards the negative direction. It indicates that the QGP medium created at lower values of $\sqrt{s_{NN}}$, which corresponds to a baryon-rich medium, has a positive Seebeck coefficient. The positive Seebeck coefficient means that the induced electric field in the medium is aligned in the same direction as the temperature gradient. For the QGP medium created at higher values of $\sqrt{s_{NN}}$, the Seebeck coefficient is negative and indicates the opposite alignment of the induced field with respect to temperature gradients in the medium. The right panel of Fig.~\ref{E0} represents the $Th_B$ as a function of $\sqrt{s_{NN}}$. Here, we can also see that for lower values of $\sqrt{s_{NN}}$, nearly up to 20 GeV, the $Th_B$ is negative. It vanishes near $\sqrt{s_{NN}} \approx 20$~GeV and then increases towards the positive direction. For a baryon-rich QGP medium, having a negative $Th_B$, which may indicate that the heat is continuously released from the hotter region to the colder region when current flows from a region of lower temperature to a higher temperature. On the contrary, with the same condition of current flow and temperature gradient, the heat is continuously absorbed at the hotter region for a QGP medium created at higher values of $\sqrt{s_{NN}}$ as it has positive values of $Th_B$.

The left panel of Fig.~\ref{Transverse} presents $NB$ as a function of $\sqrt{s_{NN}}$ for Au+Au collisions at an impact parameter of $b = 7.2$ fm. The results are shown for different magnetic field decay times, $\tau_B = 3.0$ fm and $\tau_B = 7.0$ fm, along with a comparison with the zero magnetic field case ($eB = 0.0$). The coefficient $NB$, which quantifies the generation of an electric field perpendicular to both the magnetic field and temperature gradient, is found to decrease with collision energy and saturate at higher $\sqrt{s_{NN}}$. We observe that for the lower values of $\sqrt{s_{NN}}$, nearly up to 20 GeV, there is a sharp fall in values of $NB$. As expected, no Nernst effect appears in the absence of a magnetic field. This behavior of $NB$ also reflects the parameterized behavior of $\mu_B$ as a function of $\sqrt{s_{NN}}$. For lower values of $\sqrt{s_{NN}}$, the $\mu_B$ is high, but it decreases with increasing $\sqrt{s_{NN}}$. A higher number of baryons over the anti-baryons leads to significant Hall-like effects in the presence of a magnetic field. On the other hand, for an equal number of baryons and anti-baryons, the Hall-like effects get canceled due to equal and opposite Lorentz force. The sensitivity of the $NB$ to both collision energy and the lifetime of the early-time magnetic field provides valuable insights into the transport properties of the magnetized QGP. The right panel of Fig.~\ref{Transverse} presents $Th_N$ as a function of the $\sqrt{s_{NN}}$ for the same parameters as the left panel. Here, we observe that the values of $Th_N$ are decreasing in magnitude as $\sqrt{s_{NN}}$ increases. Below the values of $\sqrt{s_{NN}}$ = 20 GeV, this fall is rapid, but for the higher values of $\sqrt{s_{NN}}$, this fall is slower. Nearly, at $\sqrt{s_{NN}}$ = 200 GeV the coefficient $Th_N$ approaches zero. 
\section{Summary}\label{sec-summary}
In summary, for the first time, we estimate the Thomson coefficient in the QGP medium. We use the kinetic theory formalism to solve the Boltzmann transport equation based on the RTA approach to calculate thermoelectric coefficients. The Thomson coefficient quantifies the heat generation or absorption in the medium when current flows in the medium in the presence of a temperature gradient. We have extended the formalism to calculate the thermoelectric coefficient considering the presence of a time-varying magnetic field. The particle's energy gets quantized via the Landau quantization in the presence of the magnetic field, but for the sake of now, the same effect is not considered in this study. It is also to be noted that heat conduction in any medium demands a conserved charge. Here, for the QGP medium, we have baryon number conservation. For this case, a non-zero baryon chemical potential is necessary. Otherwise, thermoelectric coefficients diverge at zero baryon chemical potential. For numerical estimation, we use a quasiparticle model that reproduces the lattice QCD EoS of QGP. In the presence of non-zero thermoelectric coefficients, the electric current and heat current get modified. The presence of a non-zero Seebeck coefficient also leads to a reduction in the thermal conductivity of the medium as $\kappa_0$ reduces to $\kappa_0-T\sigma_{el}S^2$. For a medium, if the Seebeck coefficient is independent of the temperature, the Thomson coefficient vanishes for it. The temperature dependence of the Seebeck coefficient is studied widely in the literature for QGP medium~\cite{Singh:2024emy, Das:2021qii1} and hadronic medium~\cite{Das:2020beh, Singh:2025rwc,Harutyunyan:2016rxm}. In our current study, we observe that the Thomson coefficient is positive for lower values of $\mu_B$, and it increases with an increase in temperature. On the other hand, the Thomson coefficient turns negative for the higher values $\mu_B$. 
The magnetic field significantly lowers the Thomson coefficient at lower values of $\mu_B$. For higher values of $\mu_B$, both Thomson and magneto-Thomson coefficients seem to saturate at higher values of temperature. In this study, we also analyze the center of mass energy $\sqrt{s_{NN}}$ dependence of the Thomson coefficient. We observe that the coefficient is negative for lower values of $\sqrt{s_{NN}}$, whereas for higher values of $\sqrt{s_{NN}}$, the coefficient is positive. The presence of non-zero Thomson coefficients in a medium affects the heat propagation in the medium. Hence, in the context of heavy-ion collisions, the presence of the Thomson coefficient may reduce or enhance the cooling of the QGP medium during the evolution. Due to Hall-like effects in the QGP medium in the presence of a magnetic field, the Nernst effect originates, which describes the transverse voltage production in the medium. The non-zero Nernst coefficient $NB$ of the medium further introduces the transverse Thomson coefficient $Th_N$. In the presence of a magnetic field, the coefficient $Th_N$ affects the temperature distribution in the medium. Both $NB$ and $Th_N$ vanish in the absence of a magnetic field.

In material science, the Thomson effect helps characterize the thermoelectric response of materials, offering insights into electron and phonon interactions. By understanding how heat and electrical currents interact in different materials, researchers can design advanced thermoelectric materials for energy harvesting, solid-state cooling, and waste heat recovery~\cite{starkov2018magnetic}. In spintronics, where electron spin and charge are manipulated for information processing, the Thomson coefficient becomes essential in understanding spin-dependent thermoelectric effects. Spin-caloritronics, a subfield of spintronics, explores how temperature gradients influence spin currents and magnetization dynamics~\cite{STARKOV2020165949}. In Ref.~\cite{Liu:2020dxg}, it is also discussed that a non-zero spin current can be produced due to the presence of the spin Hall effect. In analogy to the Hall current associated with an electric field, here, a spin current traverses in the direction of the electric field. The presence of an induced electric field~\cite{Singh:2024emy} because of non-zero thermoelectric coefficients may also further contribute to the spin Hall effect. Recently, a detailed observation of the transverse Thomson effect has been studied for semimetallic alloys in Ref.~\cite{takahagi}. The phenomenon of spontaneous magnetization in the QGP medium can also lead to an anomalous transverse Thomson effect, which is a remaining task for future work. Therefore, a deeper understanding of thermoelectric effects in QGP medium is crucial to studying the spin-relevant effects, such as the spin Hall effect, spin polarization~\cite{Singha:2022rlu}, etc.  This work is the first step to explore the higher-order thermoelectric phenomenon in the QGP medium in the context of heavy-ion collisions.  
\acknowledgments
K.S. acknowledges the doctoral fellowship from the UGC, Government of India. R.S. gratefully acknowledges the DAE-DST, Govt. of India funding under the mega-science project – “Indian participation in the ALICE experiment at CERN” bearing Project No. SR/MF/PS-02/2021-IITI (E-37123). The authors would like to thank Kshitish K. Pradhan for careful reading of the manuscript and making valuable suggestions.

\end{document}